\documentclass[abbrv,aps,prl,twocolumn,showpacs,preprintnumbers,amsmath,amssymb,
superscriptaddress,nofootinbib]{revtex4}

\usepackage{amsmath,amssymb,amsfonts}
\usepackage{bm,graphicx,color}

\newcommand{\laco}[1]{$\mathrm{LaCoO_3 }$}
\newcommand{\ket}[1]{$( #1 )$}

\begin{document}
\title{Disproportionation and Metallization at Low-Spin to High-Spin Transition in Multiorbital Mott Systems}
\author{Jan Kune\v{s} and Vlastimil K\v{r}\'apek}
\affiliation{Institute of Physics, Academy of Sciences of the Czech republic, Cukrovarnick\'a 10,
Praha 6, 162 53, Czech Republic}
\date{\today}

\begin{abstract}
We study the thermally driven spin state transition in a two-orbital Hubbard model
with crystal field splitting, which provides
a minimal description of the physics of \laco{}. We employ the dynamical mean-field theory 
with quantum Monte-Carlo impurity solver. 
At intermediate temperatures
we find a spin disproportionated phase characterized by checkerboard order of sites with
small and large spin moments. The high temperature transition from the disproportionated to a homogeneous
phase is accompanied by vanishing of the charge gap. With the increasing crystal-field
splitting the temperature range of the disproportionated phase shrinks and eventually
disappears completely.

\end{abstract}
\maketitle
The pressure or thermally driven spin state transitions play an important role in the physics of magnetic oxides \cite{gutl04}.
A notorious example is \laco{}. Its peculiar magnetic and transport properties
have attracted attention of physicists for decades, yet interpretation
of its behavior remains controversial. 
The main characteristics of \laco{} are the temperature ($T$)
dependencies of the magnetic susceptibility and the conductivity~\cite{PhysRev.155.932,PhysRevB.50.3025}, which exhibit three distinct
regions: (i) a low-$T$ non-magnetic insulator, (ii) an intermediate-$T$ paramagnetic insulator
and (iii) a high-$T$ paramagnetic bad metal. It is generally believed that the evolution from
the non-magnetic to the paramagnetic state is due to thermal population of an excited state
of the Co ion with spin (or spin and orbital) degeneracy.
Commonly considered scenarios involve either low-spin to high-spin excitation~\cite{Heikes19641600,naiman:1044,PhysRev.155.932,PhysRevLett.97.247208},
low-spin to intermediate-spin excitation~\cite{PhysRevB.54.5309} or both~\cite{PhysRevB.66.094408,JPSJ.67.290}. Temperature affects the electronic system also indirectly, by
changing the crystal-field splitting through the lattice thermal expansion.
Yet another possible piece to the puzzle are deformations of CoO$_6$ octahedra and 
their coupling to the spin states of the Co ion~\cite{PhysRev.155.932,PhysRevB.66.094408}.
Therefore it is rather difficult to distinguish the leading effects from the secondary ones.

In this Letter we study a minimal fermionic lattice model that exhibits the spin state transition.
As purely electronic it does not include the effect of lattice thermal expansion
or the magneto-elastic coupling. The model is essentially the same as the one
studied by Werner and Millis~\cite{PhysRevLett.99.126405}, who used the dynamical mean-field theory~\cite{RevModPhys.68.13} (DMFT) to map out
its phase diagram at fixed temperature, and by Suzuki {\it et al.}~\cite{PhysRevB.80.054410}, who
studied its ground state as a function of doping by variational Monte-Carlo.
We employ the DMFT method to study the temperature
dependent properties in the vicinity of the boundary between the low-spin and high-spin phases.
Unlike Ref.~\onlinecite{PhysRevLett.99.126405} we assume a specific lattice, which allows us to investigate
the ordering tendencies.
We compute~\cite{Albuquerque20071187} the one-particle spectra and the local as well as the uniform spin susceptibility and find
that, similar to the behavior of \laco{}, the model exhibits three distinct temperature regions.
In addition to the low-$T$ non-magnetic insulator and high-$T$ local moment metal 
we find a spin disproportionated insulating phase at intermediate temperatures.
In order to interpret the DMFT results we construct an effective
low-energy Potts model, which allows analytic calculations.

Our starting point is a Hubbard Hamiltonian on a square lattice
\begin{equation}
\label{eq:hubbard}
\begin{split}
H=&\sum_{i,\sigma}\bigl((\Delta-\mu) n^a_{i,\sigma}-\mu n^b_{i,\sigma}\bigr)+ 
  \sum_{\langle ij\rangle,\sigma}\bigl(t_{aa} a_{i,\sigma}^{\dagger}a_{j,\sigma}+t_{bb} b_{i,\sigma}^{\dagger}b_{i,\sigma}\bigr) \\
&+  U\sum_i \bigl(n^a_{i,\uparrow}n^a_{i,\downarrow}+n^b_{i,\uparrow}n^b_{i,\downarrow}\bigr)+
  (U-2J)\sum_{i,\sigma} n^a_{i,\sigma}n^b_{i,-\sigma}\\
&+(U-3J)\sum_{i,\sigma} n^a_{i,\sigma}n^b_{i,\sigma},
\end{split}
\end{equation}
where $a_{i,\sigma}^{\dagger}$, $b_{i,\sigma}^{\dagger}$ ($a_{i,\sigma}$, $b_{i,\sigma}$) 
are the fermionic creation (annihilation) operators for a spin index $\sigma$ and two types
of orbitals for each lattice site $i$, and $n^a_{i,\sigma}$, $n^b_{i,\sigma}$ are the corresponding occupation number operators.
The nearest-neighbor hoppings $t_{aa}=0.45$~eV, $t_{bb}=0.05$~eV and the on-site interaction parameters $U=4$~eV and $J=1$~eV
are chosen to yield a broad $a$-band and a narrow $b$-band, mimicking the electronic structure of \laco{}.
The crystal-field splitting is denoted with $\Delta$.
The $T$-dependent chemical potential $\mu$ is fixed to yield the average filling of 
2 electrons per lattice site. Unlike Ref.~\onlinecite{PhysRevLett.99.126405} we consider only Ising terms in the on-site
interaction. 
\begin{figure}
\includegraphics[width=0.8\columnwidth,clip]{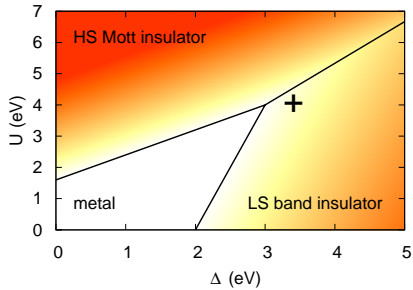}
\caption{\label{fig:phase} (Color online) Conceptual phase diagram of the two band model for $U/J=4$. The size of the 
charge gap is indicated by color intensity (white=no gap). The parameters of the present study
are marked with the cross.}
\end{figure}
  
The context of the present study is set by a conceptual form~\cite{epjst_180_5} of the $U-\Delta$ 
phase diagram of Ref.~\onlinecite{PhysRevLett.99.126405} shown in Fig.~\ref{fig:phase}. The boundary of 
the metallic phase is given by opening of a linearly increasing charge gap (indicated by the color intensity). The 
line separating high-spin (HS) Mott and low-spin (LS) band insulator corresponds to degeneracy of local HS and LS states.
The parameter range of interest corresponding to small gap LS insulator
is close to the triple point. 
In the following we present the results for $\Delta$ of 3.40 and 3.42~eV (marked with the cross in Fig.~\ref{fig:phase}).

\begin{figure}
\includegraphics[width=0.8\columnwidth,angle=270,clip]{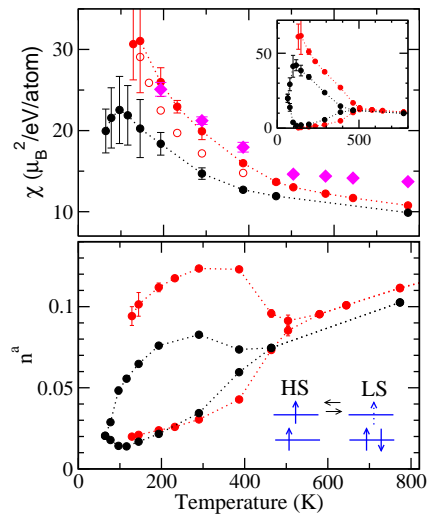}
\caption{\label{fig:disp} (Color online) The upper panel shows the $T$ dependence of the average local susceptibility per atom for
the crystal field parameter $\Delta$ of 3.40 (red) and 3.42 (black) eV. 
(The site resolved contributions are shown in the inset.) The data for $\Delta=3.40$~eV are compared
to the uniform susceptibility (diamonds) and the local susceptibility obtained in the homogeneous phase (empty circles).
The lower panel shows the difference $\bar{n}^a_A-\bar{n}^a_B$ of the occupancies between the $A$ and $B$ sites.
The dotted lines are guides to the eye.}
\end{figure}

The DMFT equations are solved as described in Ref.~\onlinecite{kunes-nio} using the strong coupling 
continuous time quantum Monte-Carlo solver~\cite{PhysRevLett.97.076405}. Guided by
proposals of spin disproportionation in \laco{}~\cite{PhysRev.155.932,PhysRevB.79.014430} we have doubled the unit cell 
to allow for a spontaneous two-sublattice order (the sublattices are denoted with $A$ and $B$).
In the lower panel of Fig.~\ref{fig:disp} we show the occupation numbers $\bar{n}^a_{A,B}$
as a function of temperature. Below 500~K a disproportionation takes place.
The local states on the $A$ sites can be described as HS+LS statistical mixture 
of LS \ket{a^0b^2} and HS \ket{a^1b^1} configurations with short excursions to 1-electron \ket{a^0b^1} configuration. 
The $B$ sites host the LS \ket{a^0b^2} states which experience short excursions
to 3-electron \ket{a^1b^2} configurations. The  
driving force of the disproportionation is a superexchange interaction 
mediated by these excursions, as shown in the inset of
the lower panel of Fig.~\ref{fig:disp}.

In the upper panel of Fig.~\ref{fig:disp} we show the local susceptibility averaged over
$A$ and $B$ sites (the $A$ and $B$ contributions are shown in the inset).
The local susceptibilities reflect the probability to find a given site in the HS state,
while the occupancies $\bar{n}^a_{A,B}$ are to some extent affected by the charge 
fluctuations to 1- and 3-electrons states on $A$ and $B$ sites respectively.
Comparison to the local susceptibility calculated in the (by constraint)
homogeneous phase reveals an enhancement of the average HS abundance due to the disproportionation.
We have calculated also the uniform susceptibility by adding a small Zeeman field to Eq.~\ref{eq:hubbard}.
In the disproportionated phase the uniform susceptibility coincides with 
the averaged local susceptibility. This is a simple consequence of 
the local moments on $A$ sites being separated from each other by the $B$ sites hosting the LS singlets.
In the high-$T$ homogeneous phase the uniform susceptibility is found to be enhanced over its
local counterpart. This is somewhat counterintuitive since a naive expectation
of an anti-ferromagnetic superexchange between the neighboring HS excitations should lead  
to an opposite effect. The Pauli susceptibility associated with the bad metallic state, discussed next, is an order
of magnitude too small to provide an explanation. Therefore we conclude that an effective ferromagnetic coupling
exists in the high-$T$ phase.

\begin{figure}
\includegraphics[width=0.7\columnwidth,angle=270,clip]{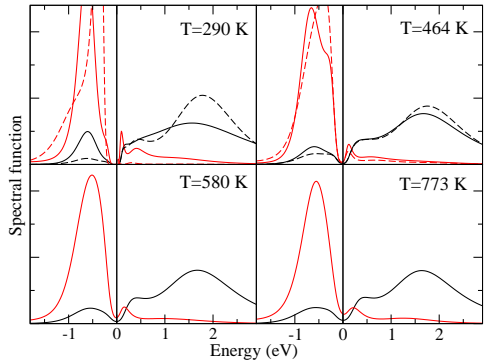}
\caption{\label{fig:spec} (Color online) The $T$ evolution the one-particle spectra for $\Delta=3.40$.
The spectral densities of different orbitals are resolved by color: $a$ (black) and $b$ (red),
while the dashed lines correspond to the $B$ sites and full lines to the $A$ sites.} 
\end{figure}

The evolution of the one-particle spectra is shown in
Fig.~\ref{fig:spec}. The disproportionated phase exhibits a well-defined charge gap which starts to fill 
with incoherent excitations as the system approaches the transition to the homogeneous phase. To
interpret this behavior we consider the definition of the charge gap  in terms
of the eigenenergies of the system $E_{N+1}+E_{N-1}-2E_N$, where $E_N$ corresponds to an eigenstate with 
a non-vanishing occupancy and $E_{N+1}$, $E_{N-1}$ are energies of lowest states that can be reached
by adding or removing an electron. 
At zero temperature the ground state is a product of LS \ket{a^0b^2} configurations
on each site. The lowest $N+1$ state corresponds to a single $a$-electron propagating over the
the filled $b$ band, while the lowest $N-1$ state corresponds to a single $b$-hole. 
The gap is obtained as the on-site contribution reduced by half-bandwidths 
$U-5J+\Delta-W_a/2-W_b/2$. 

The situation at elevated temperatures is more complicated as the initial states
involve also sites in HS configuration as well as sites with one or three electrons.
The disproportionated and the homogeneous phases differ
in the constraints posed on the relaxation of $N+1$ and $N-1$ excitations 
by the states of the neighboring sites. While in the homogeneous
phase each site has neighbors which fluctuate into 1- and 3-electron states,
in the disproportionated phase fluctuations are either to 1-electron states on $A$ sites
or 3-electron states on $B$ sites. As an example we discuss the 
$b$-electron excitations on an $A$ site from the HS initial configuration. 
(The LS initial configurations contribute with completely filled valence band.)
In the disproportionated phase the energy of lowest $N+1$ state reached by adding $b$ electron 
consists of the on-site contribution reduced by $W_a/2$ due to $a$ electron hopping.
The lowest $N-1$ state reached by $b$ electron removal corresponds to \ket{a^1b^1} configuration on the $A$ site and \ket{a^0b^1}
configuration on the neighboring $B$ site. The corresponding gap estimate is still finite \ket{U-2J-W_a/2}. 
In the homogeneous phase the lowest $N-1$ excitation connected with $b$ electron removal is the same
as just mentioned. For $N+1$ excitations there is an additional possibility in the homogeneous phase
to add $b$ electron on a site in HS state while its neighbor is in \ket{a^0b^1} configuration leading
to the \ket{a^0b^2} final state with \ket{a^1b^1} on the neighboring site after $a$ electron transfer. 
Considering these excitations we obtain a vanishing estimate for the charge gap.

\begin{figure}
\begin{center}
\includegraphics[width=0.8\columnwidth,clip]{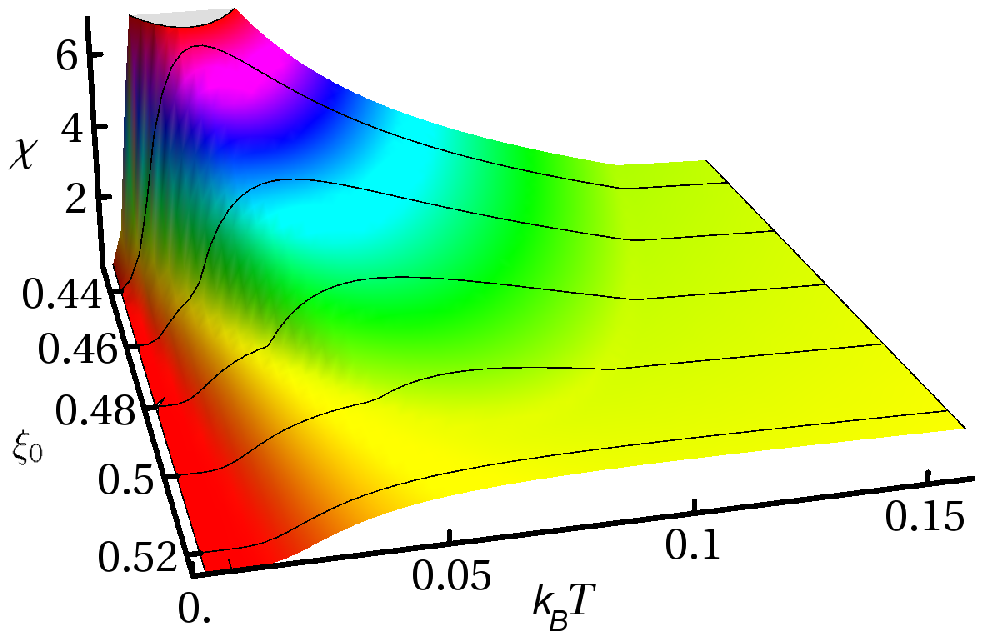}
\includegraphics[width=0.8\columnwidth,clip]{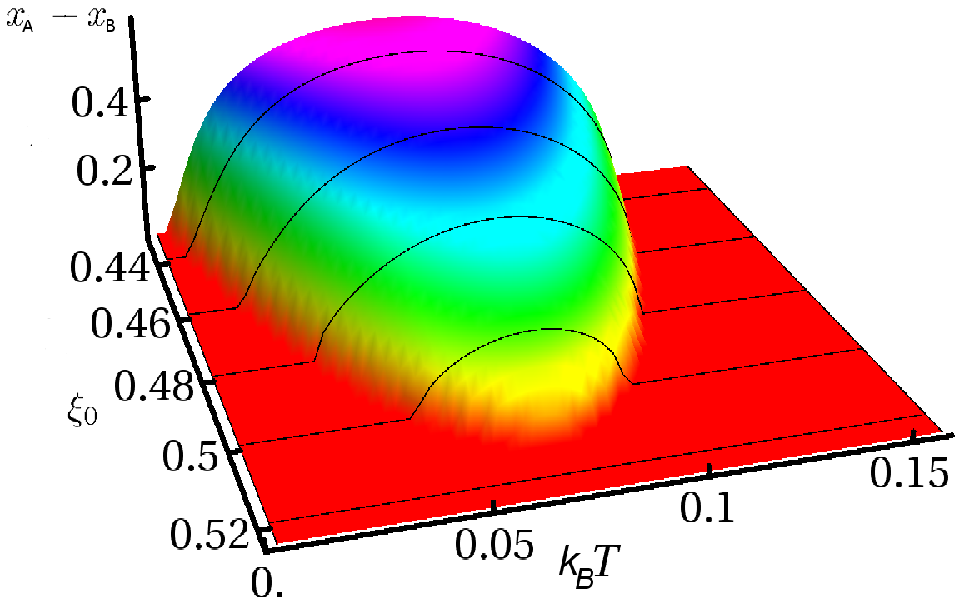}
\end{center}
\caption{\label{fig:potts} (Color online) The uniform susceptibility of the Potts model as a function
of temperature and the parameter $\xi_0$. (upper panel) The difference
of the HS populations $x_A-x_B$ as a function of temperature and $\xi_0$. (lower panel)}
\end{figure}

In order to gain insight into the DMFT results we integrate out 
charge fluctuations in (\ref{eq:hubbard}) to get an effective Potts model with three lowest-energy states 
(LS, HS$\uparrow$, HS$\downarrow$) per site
\begin{equation}
\label{eq:potts}
\tilde{H}=\xi_0\sum_{i,\sigma}n^{\text{HS}}_{i,\sigma}+\sum_{\langle ij\rangle,\sigma}\bigl(\xi_1 n^{\text{LS}}_i n^{\text{HS}}_{j,\sigma}
        + \xi_2 n^{\text{HS}}_{i,\sigma}n^{\text{HS}}_{j,-\sigma}\bigr).
\end{equation}
Here $n^{\text{HS}}_{i,\sigma}$ and $n^{\text{LS}}_i$ are the projectors of the three states. The coupling constants
arising from lowest order virtual hopping processes read $\xi_0=\Delta-3J$, $\xi_1=-\tfrac{t_{aa}^2}{U-2J}$, and
$\xi_2=-\tfrac{2t_{aa}^2}{U+J}$. Is should be pointed out that this model is only good for qualitative 
comparison to the DMFT data as the charge fluctuations are not negligible (in particular
in the metallic phase).
A mean-field decoupling of (\ref{eq:potts}) allowing for a two sublattice order leads to the 
free energy per site expression
\begin{equation}
\label{eq:f}
\begin{split}
&F(T)=\frac{\xi_0}{2}(x_A+x_B)+2\xi_1\bigl(x_A+x_B-2x_Ax_B\bigr)-\xi_2x_Ax_B \\
     & +\frac{T}{2}(1-x_A)\ln(1-x_A)+\frac{T}{2}(1-x_B)\ln(1-x_B) \\
     &+\frac{T}{2}x_A\ln(\frac{x_A}{2})+
       \frac{T}{2}x_B\ln(\frac{x_B}{2}),
\end{split}
\end{equation}
where $x_{A,B}$ are the mean values of $n^{\text{HS}}_{\uparrow}+n^{\text{HS}}_{\downarrow}$ on the 
two sublattices. The equilibrium values of $x_A-x_B$ obtained by minimization of (\ref{eq:f})
are shown in Fig.~\ref{fig:potts} together with corresponding uniform spin susceptibility.
For $\xi_0>4\xi_1$ we find a uniform LS ground state at $T=0$, which is followed by a transition
into the disproportionated phase characterized by $x_A\neq x_B$ between temperatures ${T_c}_1$ and
${T_c}_2$. With increasing $\xi_0$ the ${T_c}_1$ and ${T_c}_2$ converge and the disproportionated
phase eventually disappears for large enough $\xi_0$.

The disproportionated phase exhibits an enhanced susceptibility, which has two sources. First, like in the DMFT results
the HS abundance is enhanced with respect to the homogeneous phase.
Second, in the homogeneous
phase the anti-ferromagnetic (AFM) coupling $\xi_2$ reduces the uniform susceptibility.
Note that this is in contrast to the DMFT results.

The HS-LS model is not new. It was suggested for \laco{} by Goodenough and Raccah~\cite{PhysRev.155.932}
and the disproportionation was studied by Bari and Sivardi\`ere~\cite{bari72}.
However, traditionally the interaction between the LS and HS state
was associated with the magneto-elastic coupling, namely a breathing distortion of 
the CoO$_6$ octahedra. This is quite different from the present work where the disproportionation  
is of purely electronic origin.

Finally, we discuss the implications of our results in the context of \laco{}.
We show that physics of \laco{} over the whole temperature range can be qualitatively
described by purely electronic effects and with only one local moment state. The simplicity of the present model
lends it generality, but inevitably involves approximations.
First, our model involves two non-degenerate orbitals while \laco{} is characterized by three-fold quasi-degenerate
$t_{2g}$ and two-fold degenerate $e_g$ orbitals leading to different entropic contributions. Nevertheless, the main control parameters are the coupling constants,
and different band degeneracies have only quantitative effect.
Second, the Co-O bond-lengths change due to the normal and anomalous lattice thermal expansion~\cite{PhysRevB.66.094408}. The latter is caused by populating the HS state, well-known to weaken the metal ion\,--\,ligand bonds~\cite{PhysRevLett.89.205504,PhysRevLett.102.146402}. 
Considering a path of decreasing $\xi_0(T)$ in Fig.~\ref{fig:potts} it is clear that the lattice response
enhances the observed effects as smaller crystal field favors the disproportionation
as well as the metallization.
Third, no disproportionation and corresponding breathing lattice distortion
was observed in recent experiments~\cite{PhysRevB.66.094408,PhysRevB.67.224423}.
It is well known that mean-field approximations overestimate ordering tendencies and that a long-range
order in the mean-field solution is often indicative of short-range correlations in the system.
Therefore we speculate that in \laco{} dynamical HS-LS correlations take place. It is plausible
that such a dynamical effect can arise from the instantaneous HS-LS interaction of electronic origin.
On the other hand, magneto-elastic HS-LS interaction retardation effects due to the lattice dynamics
are likely to weaken dynamical HS-LS correlation considerably.
Fourth, two regions of Curie-Weiss behavior observed in experiments have been interpreted as an evidence for two different
local moment states (high-spin and intermediate-spin)~\cite{JPSJ.67.290}. In our model we observe different behaviors of 
the uniform susceptibility
in the disproportionated and in the homogeneous high $T$ phase. The difference has three sources: (i) an enhanced abundance
of the HS configurations in the disproportionated phase (over a hypothetical homogeneous phase at the same $T$), 
(ii) absence of the nearest-neighbor anti-ferromagnetic correlations in the disproportionated phase,
(iii) the metallicity of the homogeneous phase.

In conclusion, we have used numerical DMFT method to study a two-band Hubbard model with quasi-degenerate 
high-spin and low-spin local states. Varying temperature we have observed three different regimes:
a low-spin insulator, an insulating phase with HS-LS disproportionation and enhanced Curie-Weiss susceptibility and
a homogeneous metallic phase with Curie-Weiss susceptibility. We have argued that our model study captures
the essential physics of \laco{} and thus that the properties of \laco{} can be explained
with a single magnetic moment carrying state and without the effect of the lattice thermal expansion.

We than Z. Jir\'ak and P. Nov\'ak for numerous discussions and critical reading of the manuscript. This work was supported by
the Grant No. P204/10/0284 of the Grant Agency of the Czech Republic
and by the Deutsche Forschungsgemeinschaft through FOR1346.

\bibliography{v6}

\end{document}